\documentclass[letterpaper]{article} % DO NOT CHANGE THIS
\usepackage[preprint]{aaai2027}
\usepackage[hyphens]{url}  % DO NOT CHANGE THIS
\usepackage{graphicx} % DO NOT CHANGE THIS
\usepackage{natbib}  % DO NOT CHANGE THIS AND DO NOT ADD ANY OPTIONS TO IT
\usepackage{caption} % DO NOT CHANGE THIS AND DO NOT ADD ANY OPTIONS TO IT
\usepackage{algorithm}
\usepackage{algorithmic}

\usepackage{amssymb}
\usepackage{multirow}
\usepackage{url}
\usepackage{subfigure}
\usepackage{amsmath}
\usepackage{amsfonts}
\usepackage{subcaption}
\usepackage{graphicx}
\usepackage{pifont}

\usepackage[utf8]{inputenc}
\usepackage[T1]{fontenc}
\usepackage{xcolor}
\usepackage{colortbl}
\usepackage{soul} 
\usepackage[most]{tcolorbox} 

\definecolor{c_refrow}{HTML}{F2F2F2}
\definecolor{c_group_prop}{HTML}{FCEFE6}
\definecolor{c_group_open}{HTML}{E8F1F8}
\definecolor{c_group_caption}{HTML}{F0EEF8}
\definecolor{c_group_ours}{HTML}{EAF6EE}

\definecolor{c_temporal}{HTML}{D9D9D9}   % 浅灰色 Temporal Boundary
\definecolor{c_sfx}{HTML}{D9EAD3}        % 浅绿色 Sound Effect
\definecolor{c_music}{HTML}{C9DAF8}      % 浅蓝色 Music
\definecolor{c_speech}{HTML}{F4CCCC}     % 浅粉色 Speech (Voice & Transcript)
\definecolor{c_attribute}{HTML}{FFF2CC}  % 浅黄色 Attribute
\definecolor{c_emotion}{HTML}{D0E0E3}    % 浅青色 Emotion
\definecolor{c_background}{HTML}{EAD1DC} % 浅紫色 Background & Context

\usepackage{newfloat}
\usepackage{listings}
\DeclareCaptionStyle{ruled}{labelfont=normalfont,labelsep=colon,strut=off} % DO NOT CHANGE THIS
\floatstyle{ruled}
\newfloat{listing}{tb}{lst}{}
\floatname{listing}{Listing}

\usepackage{booktabs}

\title{AudioMap: Cloze-and-Choice Reinforcement Learning for Time-Aware Dense Audio Captioning}
\author{
    Yan Rong\textsuperscript{\rm 1,\rm 2}\thanks{Work done during an internship at Kling Team},
    Fengji Ma\textsuperscript{\rm 1,\rm 2},
    Xu Li\textsuperscript{\rm 2}\thanks{Project Leader},
    Jinting Wang\textsuperscript{\rm 1},
    Chen Zhang\textsuperscript{\rm 2},
    Li Liu\textsuperscript{\rm 1}\corresponding
}
\affiliations{
    \textsuperscript{\rm 1}The Hong Kong University of Science and Technology (Guangzhou)\\
    \textsuperscript{\rm 2}Kling Team, Kuaishou Technology
}

\begin{document}

\maketitle

\begin{abstract}
Time-aware dense audio captioning (TDAC) aims to generate multiple fine-grained attributes (dense) of the audio with precise time boundaries (time-aware). Existing methods struggle to achieve these two goals and mainly rely on supervised fine-tuning, yielding sub-optimal performance. While reinforcement learning (RL) shows promise, applying it to TDAC faces two main challenges: (1) existing rewards are too coarse to supervise multi-event, multi-attribute, and multi-relation descriptions in a fine-grained manner; and (2) temporal supervision is difficult for free-form captions, where flexible event-time expressions make reliable event-time correspondence challenging.
To address these challenges, we propose \textbf{AudioMap}, a novel RL-based TDAC framework, which shifts to a unified cloze-and-choice reward paradigm.
Specifically, we introduce the Evidence Sufficiency Reward (ESR) with an asymmetric hierarchical scoring mechanism to promote fine-grained accuracy and descriptive richness across diverse acoustic dimensions. Furthermore, we design the Event-Conditioned Temporal Reward (ECTR) to structurally bind timestamps to event semantics via temporal IoU, accompanied by a dual-curriculum learning strategy to facilitate the training process.
Finally, to support this task, we construct the first time-aware fine-grained audio captioning dataset, AudioMapCap-44K, which contains 44K carefully annotated captions. 
Extensive experiments across diverse benchmarks show that AudioMap achieves state-of-the-art (SOTA) performance among open-source models and delivers competitive or superior results relative to proprietary models. Project page and release updates are available at \url{https://github.com/ryysayhi/AudioMap}.
\end{abstract}

\section{Introduction}
Time-aware dense audio captioning (TDAC) aims to generate multiple fine-grained attributes (dense) of the audio with precise time boundaries (time-aware). Unlike conventional global audio captioning~\cite{liu2024enhancing,choi2025temp4cap}, which compresses an input clip into a short holistic summary, TDAC must disentangle concurrent and successive events, characterize their fine-grained acoustic attributes, and trace how they evolve and interact over time. Such capability is fundamental for embodied audiovisual reasoning~\cite{feng2026video}, fine-grained multimedia retrieval~\cite{qin2025deep}, and downstream audio generation~\cite{rong2025audiogenie}.

\begin{figure}[t]
    \centering
    \includegraphics[width=1.00\linewidth]{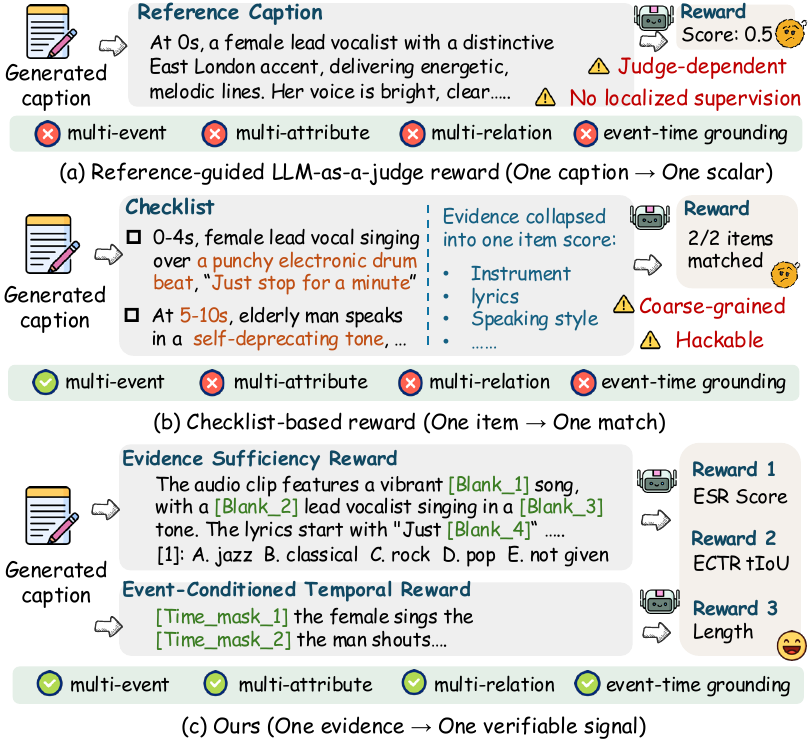}
    \vspace{-1.7em}
    \caption{Comparison of different reward paradigms.}
    \label{fig:intro}
    \vspace{-1.6em}
\end{figure}

Despite progress in audio captioning~\cite{xu2025qwen3}, current models (as summarized in Table~\ref{tab:intro_comp}) remain far from this goal. Existing systems~\cite{ma2025omni,kumar2026tac} primarily rely on supervised fine-tuning (SFT)~\cite{chensft}, which learns caption content and form but does not explicitly optimize fine-grained evidence coverage or event-time alignment. Consequently, a reliable training paradigm that jointly enforces faithful fine-grained attributes and explicit temporal grounding remains underdeveloped.

Recently, reinforcement learning (RL)~\cite{christiano2017deep,shao2025deepseekmath} has shown great potential in reasoning tasks~\cite{he2026audio}. However, extending it to open-ended time-aware dense audio captioning to simultaneously handle fine-grained attributes and temporal grounding presents two fundamental challenges: (1) \textbf{existing rewards are too coarse to supervise multi-event, multi-attribute, and multi-relation descriptions in a fine-grained manner}; and (2) \textbf{temporal supervision is difficult for free-form captions, where flexible event-time expressions make reliable event-time correspondence challenging}.

\begin{table}[t]
\centering
\caption{Capability comparison among different SOTA audio captioning methods.}
\tabcolsep=3.5pt
\vspace{-0.3em}
\resizebox{\linewidth}{!}{
\begin{tabular}{lcc}
\toprule
\multirow{2}{*}{\textbf{Methods}} & \textbf{Fine-grained} & \textbf{Temporal} \\
& \textbf{Attributes} & \textbf{Boundary} \\
\midrule
Temp4Cap~\cite{choi2025temp4cap}& \ding{55} & \ding{55} \\
TAC~\cite{kumar2026tac}&  \ding{55} & \ding{51} \\
Qwen3-Omni-Captioner~\cite{xu2025qwen3}&  \ding{51} & \ding{55} \\
Omni-Captioner~\cite{ma2025omni}&  \ding{51} & \ding{55} \\
\midrule
AudioMap (Ours) &  \ding{51} & \ding{51} \\
\bottomrule
\vspace{-3em}
\end{tabular}
}
\label{tab:intro_comp}
\end{table}

\textbf{For the first challenge}, evaluating free-form captions with dense acoustic semantics is inherently subjective. Reference-guided LLM-as-a-judge rewards (Figure~\ref{fig:intro}(a))~\cite{gunjal2025rubrics} compress the entire caption into a single scalar. Because this score inherits judge-specific preferences, it may steer optimization toward favored response styles rather than better-grounded descriptions. More importantly, this scalar does not identify which acoustic evidence is missing or incorrect. Its value may also vary across repeated evaluations, leading to inconsistent supervision.
Checklist-style rewards (Figure~\ref{fig:intro}(b))~\cite{chenavocado} remain too coarse for dense audio because their context-agnostic items can entangle overlapping sources and overlook fine-grained attributes or relations. Moreover, optimizing the number of matched items can introduce shortcut behaviors, such as repeating checklist-related keywords or producing unnecessarily long unsupported descriptions.

\textbf{For the second challenge}, existing temporally aligned benchmarks and evaluation protocols typically assume explicit associations between event descriptions and temporal segments~\cite{primus2025tacos,kumar2026tac}, which free-form captions do not provide. The evaluator must instead recover and align events and timestamps from the generated text. 
% This becomes particularly challenging for concurrent or similar events, which may be expressed differently in free-form captions. 
% This becomes particularly challenging when concurrent or similar events have nearby or overlapping temporal spans, as temporal proximity alone cannot establish event correspondence.
This becomes particularly challenging for concurrent or similar events, where temporal proximity alone cannot establish event correspondence.
Event-agnostic matching based on isolated numeric tokens or global time overlap can therefore assign temporal credit to a plausible interval even when it is associated with the wrong semantic event.

To address these challenges, we propose \textbf{AudioMap}, a novel RL-based TDAC framework that jointly optimizes faithful fine-grained attributes and explicit event-time grounding. The core of our approach lies in a fundamental paradigm shift: we formulate comprehensive caption evaluation as a rigorous unified cloze-and-choice reward paradigm. \textit{A caption is considered high-quality if it provides sufficient grounded evidence to pass a diverse set of fine-grained factual examinations, and explicitly grounds each described semantic event to its precise temporal boundaries.}

More precisely, for the first challenge, instead of using coarse and hackable checklist matching, we propose the \textbf{E}vidence \textbf{S}ufficiency \textbf{R}eward (\textbf{ESR}). ESR systematically generates a collection of unpredictable, multi-dimensional multiple-choice cloze questions covering diverse fine-grained acoustic dimensions. 
By decomposing holistic assessment into localized examinations, ESR provides fine-grained supervision over source-grounded facts, avoiding ambiguous scalar feedback and coarse checklist supervision.
We further introduce asymmetric hierarchical scoring, penalizing incorrect evidence more strongly than missing evidence to encourage dense yet faithful descriptions.

For the second challenge, instead of rewarding timestamps in isolation or matching rigid, token-level strings, we propose the \textbf{E}vent-\textbf{C}onditioned \textbf{T}emporal \textbf{R}eward (\textbf{ECTR}). 
ECTR formulates temporal evaluation as an event-conditioned cloze test: given an event anchor, the judge extracts its predicted interval and compares it with the reference using temporal IoU (tIoU).
To further stabilize the joint semantic-temporal learning, we design a dual-curriculum strategy over both input duration and reward complexity.

In addition, we construct AudioMapCap-44K, the first time-aware fine-grained audio captioning dataset comprising 43,870 audio-caption pairs spanning 769.7 hours. Furthermore, beyond audio-only inputs, we extend AudioMap to support audio-visual inputs, leveraging the visual stream as an auxiliary input to the audio-centered task.

In summary, the main contributions of this work are as follows:

\begin{itemize}
\item We propose AudioMap, a novel RL-based framework for time-aware dense audio captioning, which shifts to a unified cloze-and-choice reward paradigm. To the best of our knowledge, this is the first framework that simultaneously enforces faithful fine-grained attributes and explicit event-time grounding.
\item We introduce the Evidence Sufficiency Reward, equipped with an asymmetric hierarchical scoring mechanism, to encourage descriptive richness without sacrificing faithfulness.
\item We propose the Event-Conditioned Temporal Reward, which translates temporal evaluation into a semantic cloze test optimized via tIoU, structurally binding time to text semantics. We also design a dual-curriculum learning strategy to facilitate the training process.
\item We construct the first time-aware fine-grained audio captioning dataset, AudioMapCap-44K. AudioMap achieves SOTA performance among open-source models and delivers competitive or superior results relative to proprietary models.
\end{itemize}

\begin{figure*}[t]
    \centering
    \includegraphics[width=1\linewidth]{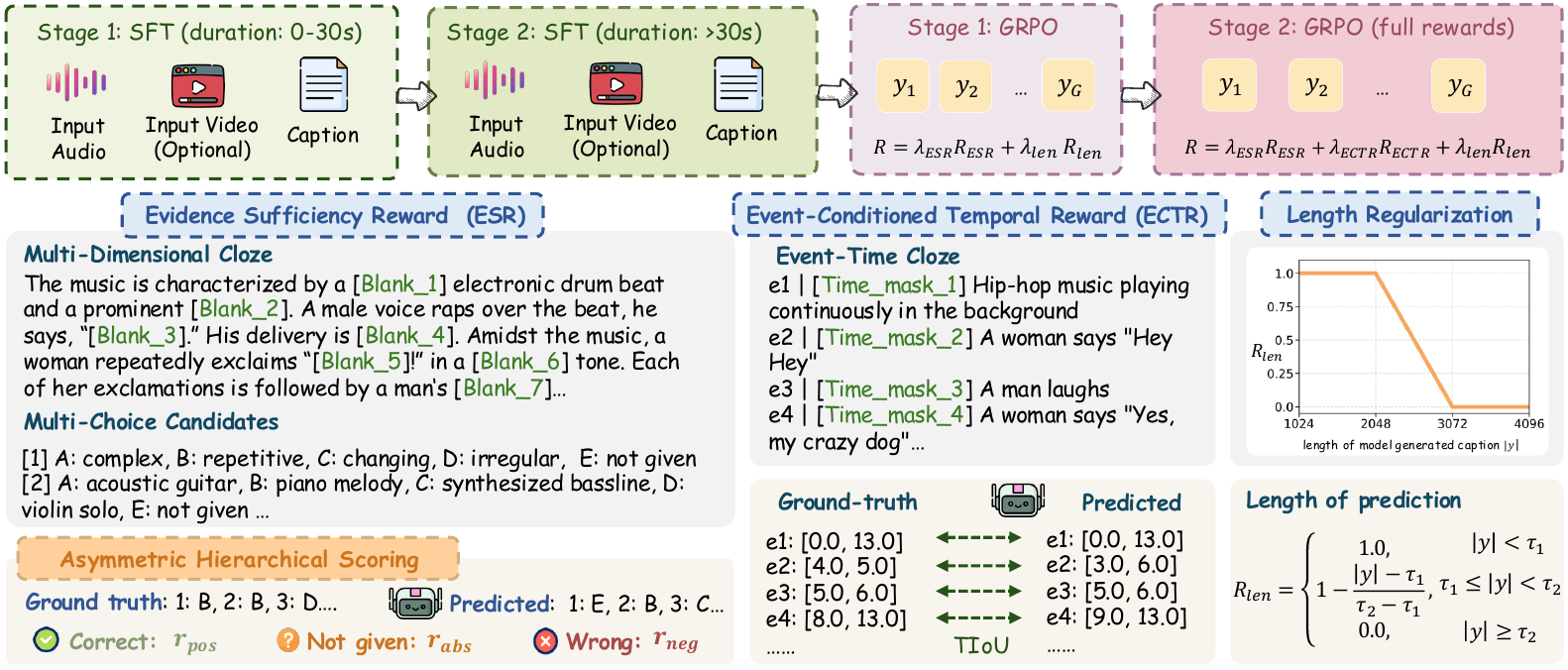}
    \vspace{-1.5em}
    \caption{The overall framework of AudioMap. SFT proceeds from shorter to longer inputs. GRPO first optimizes ESR with length regularization and then adds ECTR. The lower panels illustrate the construction and scoring of the three reward terms.}
    \label{fig:overview}
    \vspace{-0.8em}
\end{figure*}
\section{Related Work}
\subsection{Audio Captioning}
Automated audio captioning is typically formulated as a cross-modal generation task that maps input audio to natural language descriptions. Early works focused on global clip-level summaries~\cite{choi2025temp4cap}. Recently, Qwen3-Omni-Captioner~\cite{xu2025qwen3} and Omni-Captioner~\cite{ma2025omni} pursue more detailed audio descriptions. However, these methods aim to improve global quality but lack the ability to ground event semantics with precise timestamps. As a recent exception, TAC~\cite{kumar2026tac} pioneers generating structured, timestamped descriptions for audio events. Nevertheless, it remains limited in capturing fine-grained attributes.
Overall, existing methods tend to optimize either detail richness or temporal localization in isolation. In contrast, our method generates comprehensive descriptions with fine-grained attributes and precise timestamps.

\subsection{Reinforcement Learning for MLLMs}
Reinforcement learning~\cite{christiano2017deep} has become an important paradigm in multimodal understanding tasks. Video-R1~\cite{feng2026video} adopts GRPO~\cite{shao2025deepseekmath} with rule-based rewards to improve video understanding. Audio-DeepThinker~\cite{he2026audio} designs a progressive two-stage training paradigm to foster high-quality chain-of-thought emergence in audio reasoning. However, these task-specific approaches are ill-suited for dense audio captioning, as objectively verifying long descriptions remains challenging. Currently, few RL-based methods explicitly target captioning tasks. VideoChat-R1~\cite{li2025videochat} leverages event-recall rewards to improve caption quality, and AVoCaDO~\cite{chenavocado} further enhances video captioning by combining dialogue and checklist rewards. Nevertheless, existing reward designs mostly rely on global matching or coarse-grained checklists. There is still a lack of a unified training paradigm that simultaneously enforces event density, factual sufficiency, and structural event-time binding.
\section{Method}
AudioMap addresses time-aware dense audio captioning with a progressive post-training pipeline designed around two complementary goals: fine-grained semantic coverage and explicit event-time grounding. Figure~\ref{fig:overview} summarizes how supervised learning, reward-guided policy optimization, and the dual curriculum are integrated.

\subsection{Task Formulation and Training Overview}
Given an audio clip $\mathcal{A}$, AudioMap generates a time-aware dense caption $\mathcal{Y}=(y_1,\ldots,y_T)$ that describes audible events, their fine-grained attributes and relations, and their approximate timing in chronological order.

\paragraph{Supervised Fine-Tuning.}
We initialize AudioMap from Qwen2.5-Omni~\cite{xu2025qwen2_5omni} and fine-tune it on AudioMapCap-44K using both audio-only and paired audio-visual examples. 
With autoregressive next-token prediction against the reference captions, SFT teaches the model to follow the dense-caption instruction, cover audible content, and express event-level attributes with explicit time references.
% With next-token prediction on reference captions, SFT teaches the model to cover audible content and express fine-grained attributes with explicit time references.

\paragraph{Group Relative Policy Optimization.}
Starting from the SFT policy, we apply GRPO~\cite{shao2025deepseekmath} with the rewards introduced below. GRPO uses audio-only inputs. For each audio clip $\mathcal{A}$, the old policy $\pi_{\theta_{\mathrm{old}}}$ samples a group of $G$ captions $\{\mathcal{Y}_i\}_{i=1}^{G}$, and each caption receives a composite reward $R_i=R(\mathcal{Y}_i)$. The optimization objective is
\vspace{-0.1em}
\begin{equation}\small
\begin{aligned}
\mathcal{J}_{\mathrm{GRPO}}(\theta)
&=\mathbb{E}\!\left[
\frac{1}{G}\sum_{i=1}^{G}\frac{1}{T_i}\sum_{t=1}^{T_i}
\left(\ell_{i,t}^{\mathrm{clip}}-\beta D_{i,t}^{\mathrm{KL}}\right)
\right],\\
\ell_{i,t}^{\mathrm{clip}}
&=\min\!\Bigl(
\rho_{i,t}\hat{A}_i,\,
\operatorname{clip}(\rho_{i,t},1-\epsilon,1+\epsilon)\hat{A}_i
\Bigr).
\end{aligned}
\end{equation}
where $\hat{A}_i=(R_i-\bar{R})/(s_R+\delta)$ is the group-normalized advantage, $T_i$ is the number of tokens in $\mathcal{Y}_i$, and $\rho_{i,t}$ is the token-level probability ratio between $\pi_{\theta}$ and $\pi_{\theta_{\mathrm{old}}}$. $\bar{R}$ and $s_R$ denote the mean and standard deviation of rewards within the group. Here, $\delta$ prevents division by zero, $\epsilon$ is the clipping range, and $D_{i,t}^{\mathrm{KL}}$ measures the token-level divergence from the frozen reference policy $\pi_{\mathrm{ref}}$. The coefficient $\beta$ controls the KL strength. The composite reward combines evidence sufficiency, event-conditioned temporal alignment, and length regularization according to the curriculum described below.

\subsection{Evidence Sufficiency Reward}
Holistic caption scores do not reveal which events, attributes, or relations are correctly covered, omitted, or contradicted. We formulate the Evidence Sufficiency Reward (ESR) as a set of localized multiple-choice cloze tests. A caption receives a high ESR only when it contains enough evidence to recover diverse facts supported by the source audio.

\paragraph{Multi-Dimensional Cloze Construction.}
For each training sample, we construct a factual cloze set $\mathcal{Q}_{f}=\{q_k\}_{k=1}^{K}$. Each question $q_k=(p_k,\mathcal{C}_k,c_k^{*})$ contains a masked passage $p_k$, a candidate set $\mathcal{C}_k$, and a correct answer $c_k^{*}$. Rather than using this format only for caption evaluation, as in Omni-Cloze~\cite{ma2025omni}, ESR uses individual cloze questions as localized feedback during policy optimization. The reference caption provides a draft for question construction, while the source audio determines which target facts are valid. We create blanks only for audible and verifiable details, including speech content, speaker and prosodic attributes, sound events, music, acoustic scenes, background noise, audio quality, and event relations.
Each blank has five options. A--D contain the correct answer and three plausible distractors matched in semantic type and granularity, while E is fixed as \texttt{Not given}. The correct answer is randomly placed among A--D. This construction requires the generated caption to distinguish the target fact from plausible alternatives rather than match a keyword in a coarse checklist.

\paragraph{Caption-Conditioned Examination.}
A frozen examiner $g_{\phi}$ receives the cloze task and generated caption, but not the source audio or unmasked reference caption:
\begin{equation}\small
\hat{c}_k=g_{\phi}(p_k,\mathcal{C}_k,\mathcal{Y}),
\qquad
\hat{c}_k\in\{A,B,C,D,E\}.
\end{equation}
The examiner selects A--D only when the caption explicitly states or clearly entails the corresponding fact. Otherwise, it returns E. The generated caption is thus the only evidence available for resolving each blank.

\paragraph{Asymmetric Hierarchical Scoring.}
ESR distinguishes a missing detail from an incorrect concrete claim. A correct answer receives $r_{\mathrm{pos}}$, choosing E receives the omission penalty $r_{\mathrm{abs}}$, and choosing an incorrect A--D option receives $r_{\mathrm{neg}}$:
\begin{equation}
r_k =
\begin{cases}
r_{\mathrm{pos}},  & \text{if } \hat{c}_k = c_k^{*} \\
r_{\mathrm{abs}},  & \text{if } \hat{c}_k = E \\
r_{\mathrm{neg}}, &  \text{otherwise}
\end{cases}
\end{equation}
where $r_{\mathrm{neg}}<r_{\mathrm{abs}}\leq0<r_{\mathrm{pos}}$. We use $r_{\mathrm{pos}}=1$, $r_{\mathrm{abs}}=-0.5$, and $r_{\mathrm{neg}}=-1$. 
This asymmetry discourages speculative details while favoring dense descriptions supported by acoustic evidence.
The sample-level reward is $R_{\mathrm{ESR}}=\frac{1}{K}\sum_{k=1}^{K}r_k$.
% \begin{equation}\small
% R_{\mathrm{ESR}}=\frac{1}{K}\sum_{k=1}^{K}r_k.
% \end{equation}

\subsection{Event-Conditioned Temporal Reward}
Matching timestamp strings or intervals without an event identity can assign temporal credit to the wrong sound. The Event-Conditioned Temporal Reward (ECTR) instead asks the examiner to recover a time interval for a specified semantic event and scores only that event-interval pair.

\paragraph{Event-Time Cloze Construction.}
For each sample, we construct an event-time cloze set $\mathcal{Q}_{t}=\{a_m\}_{m=1}^{M}$ from the source audio and reference annotation. Each anchor $a_m=(u_m,\tau_m^{*})$ pairs a concise event description $u_m$ with a reference interval $\tau_m^{*}=[s_m^{*},e_m^{*}]$. 
The interval is hidden from the examiner, while the event identity remains visible. Each anchor describes a semantically specific principal event instance, incorporating discriminative content or attributes when necessary to distinguish concurrent or similar events. We retain principal events and transitions and exclude low-value micro-events unless they are important to the timeline.

\paragraph{Event-Conditioned Interval Extraction.}
A frozen examiner $g_{\phi}^{t}$ receives the visible event description and generated caption and returns the interval explicitly associated with that event:
\begin{equation}\small
\hat{\tau}_m=g_{\phi}^{t}(u_m,\mathcal{Y}).
\end{equation}
If the caption provides start and end times, the examiner returns $\hat{\tau}_m=[\hat{s}_m,\hat{e}_m]$. A point timestamp $t$ is expanded to $[\max(0,t-w),t+w]$ using a predefined half-width $w$. If either the event or a usable timestamp is unsupported, the examiner returns \texttt{null}.

\paragraph{Temporal IoU Scoring.}
Missing temporal evidence receives zero credit. For a valid predicted interval, $I_m=|\tau_m^{*}\cap\hat{\tau}_m|$ and $U_m=|\tau_m^{*}\cup\hat{\tau}_m|$ denote the intersection and union lengths, respectively. The event-level tIoU score is
\begin{equation}\small
r_{t,m} =
\begin{cases}
0, & \text{if } \hat{\tau}_m=\text{\texttt{null}}\ \text{or}\ U_m\leq0, \\[2pt]
\dfrac{I_m}{U_m}, & \text{otherwise},
\end{cases}
\end{equation}
ECTR then averages the scores over all anchors as $R_{\mathrm{ECTR}}=\frac{1}{M}\sum_{m=1}^{M}r_{t,m}$. Because the interval is extracted for a specified event, temporal credit depends on both event evidence and its associated boundary.

\begin{table*}[t]
\centering
\caption{Comparison with SOTA methods on audio-only input. Gray rows denote proprietary models or open-source models with more than 8B total parameters, which are included as reference points. Best and second-best results among open-source models are highlighted in bold and underlined, respectively.}
\label{result_sota}
\vspace{-0.5em}
\resizebox{0.9\linewidth}{!}{
    \begin{tabular}{lcccccc}
    \toprule
    \textbf{Model} & \textbf{Size} & \textbf{Omni-Cloze} $\uparrow$ & \textbf{MMSU} $\uparrow$ & \textbf{MMAR} $\uparrow$ & \textbf{MMAU} $\uparrow$ & \textbf{TACOS} $\uparrow$ \\
    \midrule
    \multicolumn{7}{l}{\textbf{\textit{Proprietary models}}} \\
    \rowcolor{c_refrow} GPT-4o Audio~\cite{hurst2024gpt} & -- & 38.9 & 40.5 & 53.8 & 56.3 & 35.5 \\
    \rowcolor{c_refrow} Gemini-2.5-Flash~\cite{comanici2025gemini} & -- & 47.3 & 56.7 & 58.9 & 68.0 & 41.8 \\
    \rowcolor{c_refrow} Gemini-2.5-Pro~\cite{comanici2025gemini} & -- & 57.9 & 71.8 & 66.5 & 72.8 & 51.4 \\
    \rowcolor{c_refrow} Gemini-3.1-Pro~\cite{googledeepmind2026gemini31pro} & -- & 64.1 & 69.5 & 69.6 & 70.5 & 49.6 \\
    % Qwen3.5-Omni-Flash  & - & & & & & \\
    \rowcolor{c_refrow} Qwen3.5-Omni-Plus~\cite{team2026qwen35} & -- & 57.6 & 69.8 & 62.6 & 71.6 & 60.9 \\
    \midrule
    \addlinespace[2pt]
    \multicolumn{7}{l}{\textbf{\textit{Open-source generalist models}}} \\
    Qwen2.5-Omni~\cite{xu2025qwen2_5omni} & 3B & 24.2 & 57.2 & 50.0 & 63.3 & 37.2 \\
    Qwen2-Audio~\cite{chu2024qwen2} & 7B & 0.9 & 32.9 & 30.7 & 38.0 & 20.0 \\
    Qwen2.5-Omni~\cite{xu2025qwen2_5omni} & 7B & 28.0 & 60.7 & 50.8 & 62.8 & 45.7 \\
    Kimi-Audio~\cite{ding2025kimi} & 7B & 2.9 & 43.4 & 39.8 & 54.5 & 36.3 \\
    MiDashengLM~\cite{dinkel2025midashenglm} & 7B & 25.2 & 62.2 & 51.7 & 67.6 & 37.3 \\
    Step-Audio-2-mini~\cite{wu2025step} & 8B & 5.7 & 53.6 & 44.6 & 62.5 & 37.8 \\
    Audio Flamingo 3~\cite{ghosh2026audio} & 7B & 6.3 & 42.5 & 36.7 & 47.1 & 37.3 \\
    % Qwen3-Omni-Instruct~\cite{xu2025qwen3} &30B-A3B&  & & & & \\
    \midrule
    \addlinespace[2pt]
    \multicolumn{7}{l}{\textbf{\textit{Caption-specialized models}}} \\
    Omni-Captioner~\cite{ma2025omni} & 7B & 53.2 & -- & 59.8 & 70.0 & -- \\
    AVoCaDO~\cite{chenavocado} & 7B & 48.1 & 67.0 & 61.7 & 71.0 & 43.9 \\
    TimeChat-Captioner~\cite{yaotimechat} & 7B & 43.1 & 65.6 & 58.1 & 68.8 & 47.2 \\
    \rowcolor{c_refrow} Qwen3-Omni-Captioner~\cite{xu2025qwen3} & 30B-A3B & 56.3 & \underline{68.2} & 59.8 & 68.6 & 41.5 \\
    \midrule
    \addlinespace[2pt]
    % \multicolumn{7}{l}{\textit{AudioMap}} \\
    \textbf{AudioMap (Ours)} & 3B & \underline{63.4} & 67.8 & \textbf{63.8} & \underline{71.6} & \underline{50.2} \\
    \textbf{AudioMap (Ours)} & 7B & \textbf{64.6} & \textbf{70.2} & \underline{63.3} & \textbf{72.4} & \textbf{57.4} \\
    \bottomrule
    \end{tabular}
}
\vspace{-0.5em}
\end{table*}

\subsection{Length Regularization}
Open-ended policy optimization may favor overlong outputs or repetitive loops. We use $R_{\mathrm{len}}$ as an upper-length regularizer, following the piecewise form in~\cite{chenavocado}:
\begin{equation}\small
R_{\mathrm{len}}(\mathcal{Y}) =
\begin{cases}
1.0, & \text{if } |\mathcal{Y}| \leq \tau_1, \\
1 - \dfrac{|\mathcal{Y}| - \tau_1}{\tau_2 - \tau_1},
& \text{if } \tau_1 < |\mathcal{Y}| < \tau_2, \\
0.0, & \text{otherwise}
\end{cases}
\end{equation}
where $|\mathcal{Y}|$ is the number of generated tokens. We set $\tau_1=2048$ and $\tau_2=3072$. The reward remains constant below $\tau_1$ and decays linearly thereafter, penalizing only excessively long generations.
 
\subsection{Dual-Curriculum Learning Strategy}
Long recordings increase both the number of acoustic events and the distance over which reward must be assigned. We therefore organize training along two axes: input duration and reward complexity.

\paragraph{Input Duration Curriculum.}
Across SFT and GRPO, training begins with shorter clips and later expands to a broader duration range. The early phase presents fewer event transitions and shorter output sequences. Longer recordings with more complex and overlapping content are introduced after the model has learned to cover and temporally organize events in shorter clips.

\paragraph{Reward Complexity Curriculum.}
GRPO first optimizes semantic evidence coverage together with length control, and then adds event-conditioned temporal supervision:
\begin{equation}\small
\begin{aligned}
R^{(1)}
&=\lambda_{\mathrm{ESR}}R_{\mathrm{ESR}}
+\lambda_{\mathrm{len}}R_{\mathrm{len}}, \\
R^{(2)}
&=\lambda_{\mathrm{ESR}}R_{\mathrm{ESR}}
+\lambda_{\mathrm{ECTR}}R_{\mathrm{ECTR}}
+\lambda_{\mathrm{len}}R_{\mathrm{len}}.
\end{aligned}
\end{equation}
This ordering first strengthens caption-level evidence coverage and then introduces the harder requirement of aligning individual events with their timestamps.

\section{Experimental Results}

\begin{figure*}[t]
    \centering
    \includegraphics[width=1\linewidth]{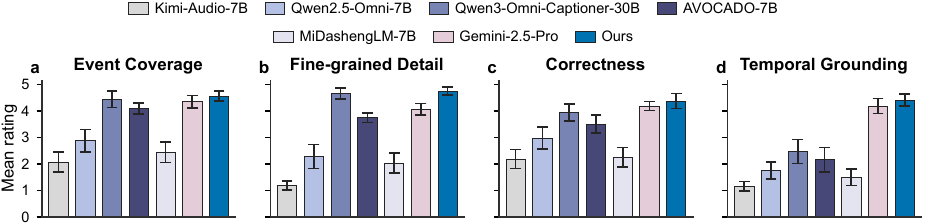}
    \vspace{-1.7em}
    \caption{Human evaluation of caption quality across four dimensions. Bars show mean ratings on the 1--5 MOS scale. Error bars denote two-sided 95\% Student's $t$ confidence intervals. AudioMap-7B is labeled as Ours.}
    \label{fig:user_study}
    \vspace{-0.8em}
\end{figure*}

% \begin{table}[t]
% \centering
% \caption{Human evaluation results across four dimensions. The values after "±" indicate standard deviations. Best performances are in bold, and second-best are underlined.}
% \label{result_user}
% \vspace{-0.5em}
% \resizebox{\linewidth}{!}{
%     \begin{tabular}{lccccc}
%     \toprule
%     \multirow{2}{*}{\textbf{Model}} & \textbf{Event} & \textbf{Fine-grained} & \multirow{2}{*}{\textbf{Correctness}} & \textbf{Temporal}\\
%     & \textbf{Coverage} & \textbf{Details} & & \textbf{Grounding} \\
%     \midrule 
%     Gemini-2.5-Pro  & $4.34\pm0.65$ & $4.06\pm0.62$ & $\underline{4.19}\pm0.47$ & $\underline{4.19}\pm0.78$ \\ 
%     \midrule
%     Qwen3-Omni-Captioner-30B & $\underline{4.43}\pm0.84$ & $\underline{4.65}\pm0.60$ & $3.94\pm0.91$ & $2.46\pm1.24$ \\
%     Kimi-AudiO-7B & $2.06\pm1.05$ & $1.19\pm0.47$ & $2.19\pm0.99$ & $1.16\pm0.51$ \\
%     MiDashengLM-7B & $2.43\pm1.07$ & $2.03\pm1.06$ & $2.25\pm1.02$ & $1.50\pm0.88$ \\ 
%     Qwen2.5-Omni-7B & $2.88\pm1.18$ & $2.28\pm1.25$ & $2.97\pm1.14$ & $1.75\pm0.92$ \\
%     AVoCaDO-7B & $4.09\pm0.58$ & $3.75\pm0.50$ & $3.50\pm0.95$ & $2.16\pm1.27$ \\ 
%     \midrule 
%     AudioMap-7B (Ours)& $\textbf{4.56}\pm0.50$ & $\textbf{4.75}\pm0.44$ & $\textbf{4.38}\pm0.79$ & $\textbf{4.41}\pm0.61$ \\
%     \bottomrule
%     \end{tabular}
% }
% \vspace{-0.3em}
% \end{table}

\subsection{Dataset}
To support the training of time-aware dense audio captioning, we construct AudioMapCap-44K. The raw audio and visual streams are sampled from the ASID-1M dataset~\cite{li2026towards}. We use Gemini-3.1-Pro~\cite{googledeepmind2026gemini31pro} in an iterative three-stage pipeline. First, it generates an initial caption that records acoustic events, temporal boundaries, and event-level attributes. 
Second, it identifies missing or underspecified information and conducts targeted follow-up questioning using the source audio, with visual evidence serving as auxiliary support for acoustically ambiguous details. Finally, it integrates the initial caption with the added information from these multi-turn responses into a cohesive, time-aware dense caption.

We apply rule-based filtering to all generated captions and use GPT-4.1~\cite{achiam2023gpt} to assess the completeness of each remaining caption. Only samples scoring 4 or 5 are retained. We then manually inspect a subset for event coverage, temporal accuracy, and hallucinations. After annotation and quality screening, the final dataset contains 43,870 high-quality caption pairs spanning 769.7 hours. The retained corpus covers diverse audio content and a broad range of clip durations. Complete dataset statistics and quality-control procedures are provided in \textbf{Appendix A}.

\begin{table}[t]
\centering
\caption{Direct caption evaluation on Omni-Cloze with audio-visual input. Formatting follows Table~\ref{result_sota}.}
\label{result_omni_av}
\vspace{-0.5em}
\resizebox{0.97\linewidth}{!}{
    \begin{tabular}{lcc}
    \toprule
    \textbf{Model} & \textbf{Size} & \textbf{Omni-Cloze} $\uparrow$ \\
    \midrule
    \multicolumn{3}{l}{\textbf{\textit{Proprietary models}}} \\
    % GPT-4o Audio~\cite{hurst2024gpt} & 38.1 \\
    \rowcolor{c_refrow} Gemini-2.5-Flash~\cite{comanici2025gemini} & -- & 47.7 \\
    \rowcolor{c_refrow} Gemini-2.5-Pro~\cite{comanici2025gemini} & -- &55.2 \\
    \rowcolor{c_refrow} Gemini-3.1-Pro~\cite{googledeepmind2026gemini31pro} & -- & 59.5 \\
    % Qwen3.5-Omni-Flash& - &  \\
    \rowcolor{c_refrow} Qwen3.5-Omni-Plus~\cite{team2026qwen35} & -- & 52.7 \\
    \midrule
    \addlinespace[2pt]
    \multicolumn{3}{l}{\textbf{\textit{Open-source generalist models}}} \\
    Qwen2.5-Omni~\cite{xu2025qwen2_5omni} & 3B & 10.5 \\
    Qwen2.5-Omni~\cite{xu2025qwen2_5omni} & 7B & 15.7 \\
    \rowcolor{c_refrow} Qwen3-Omni-Instruct~\cite{xu2025qwen3} & 30B-A3B & 12.7 \\
    \midrule
    \addlinespace[2pt]
    \multicolumn{3}{l}{\textbf{\textit{Caption-specialized models}}} \\
    Omni-Captioner~\cite{ma2025omni} & 7B & 54.5 \\
    AVoCaDO~\cite{chenavocado} & 7B & 49.4 \\
    TimeChat-Captioner~\cite{yaotimechat} & 7B & 38.9 \\
    \midrule
    \addlinespace[2pt]
    % \multicolumn{3}{l}{\textit{AudioMap}} \\
    \textbf{AudioMap (Ours)} & 3B & \underline{63.5} \\
    \textbf{AudioMap (Ours)} & 7B & \textbf{64.9} \\
    \bottomrule
    \end{tabular}
}
\vspace{-0.3em}
\end{table}

\subsection{Implementation Details}
We initialize AudioMap from Qwen2.5-Omni and use a single model for both audio-only and audio-visual inputs. SFT uses all 43,870 AudioMapCap-44K samples, with 80\% audio-only and 20\% paired audio-visual inputs, while GRPO uses 12,500 audio-only samples. The vision encoder remains frozen throughout training, while the modality aligner, LLM backbone, and language modeling head are optimized. 
Both SFT and GRPO use two-stage duration curricula, with GRPO additionally following the reward-complexity curriculum described above. Each stage is trained for one epoch.
During GRPO, Qwen3.6-27B serves as the frozen examiner for ESR and ECTR, with eight rollouts sampled per input. We use a KL coefficient of 0.06 and a cosine learning-rate schedule. Full training configurations are provided in \textbf{Appendix B}.

\subsection{Comparison with SOTA Methods}
Table~\ref{result_sota} summarizes the audio-only comparison across direct caption, QA-based, and temporal evaluations, while Table~\ref{result_omni_av} reports direct Omni-Cloze performance with audio-visual input. These three evaluation axes measure complementary aspects of time-aware dense audio captioning.

\paragraph{Direct Caption Evaluation.}
On Omni-Cloze~\cite{ma2025omni}, AudioMap-7B achieves SOTA performance in both input settings. With audio-only input, it reaches 64.6, exceeding Gemini-3.1-Pro by 0.5 points and the strongest listed open-source captioner, Qwen3-Omni-Captioner, by 8.3 points. With audio-visual input, AudioMap-7B scores 64.9, outperforming Gemini-3.1-Pro by 5.4 points and Omni-Captioner by 10.4 points. These gains show that AudioMap produces captions with stronger fine-grained evidence coverage across both audio-only and audio-visual inputs.

\begin{table}[t]
\centering
\caption{Ablation of the post-training pipeline and comparison among reward variants. Here, \textit{w/o asym. scoring} denotes ``without asymmetric scoring,'' and ``LLM-judge reward'' denotes the reference-guided LLM-as-a-judge variant. The best result in each metric is highlighted in bold.}
\vspace{-0.5em}
\tabcolsep=2.8pt
\label{tab:pipe_ablation}
\resizebox{\linewidth}{!}{
\begin{tabular}{lcccc}
\toprule
\textbf{Method}&\textbf{Omni-Cloze} $\uparrow$  & \textbf{MMAR} $\uparrow$ & \textbf{MMAU} $\uparrow$ & \textbf{TACOS} $\uparrow$ \\
\midrule
\multicolumn{5}{l}{\textbf{\textit{SFT stage}}} \\
Base Model & 28.0  & 50.8 & 62.8 & 45.7\\
AudioMap-7B-SFT & 59.7  & 60.6 & 70.6 & 54.6 \\
\quad\textit{w/o curriculum} & 59.5  & 61.5 & 68.4 & 53.3 \\
\midrule
\addlinespace[2pt]
\multicolumn{5}{l}{\textbf{\textit{GRPO stage}}} \\
AudioMap-7B-GRPO & \textbf{64.6} &  63.3 & \textbf{72.4} & 57.4 \\
\quad\textit{w/o asym. scoring} & 63.4 & 62.6 & 72.0 & 56.3 \\
\quad\textit{w/o curriculum} & 64.5 & \textbf{64.5} & 69.9 & 54.1 \\
\midrule
\addlinespace[2pt]
\multicolumn{5}{l}{\textbf{\textit{Reward variants}}} \\
LLM-judge reward & 58.7  & 62.9 & 70.2 & \textbf{58.1} \\
Checklist-based reward & 61.1 & 61.2 & 71.4 & 55.2 \\
\bottomrule
\end{tabular}
}
\vspace{-0.5em}
\end{table}

\paragraph{QA-based Caption Evaluation.}
We further assess whether the generated captions retain sufficient evidence for downstream question answering. For all three benchmarks, a text-only Qwen3.6-27B judge~\cite{qwen36_27b} answers from the caption alone, without access to the original audio. AudioMap-7B achieves 70.2 on MMSU~\cite{dingdong2026mmsu} and 72.4 on MMAU~\cite{sakshi2025mmau}, outperforming the strongest listed open-source baselines by 2.0 and 1.4 points, respectively. On MMAR~\cite{ma2026mmar}, AudioMap-3B and 7B both outperform all compared open-source models. These benchmark-specific gains indicate that AudioMap captions retain information useful for both acoustic perception and caption-conditioned reasoning.

\paragraph{Temporal Evaluation.}
AudioMap-7B achieves a TACOS~\cite{primus2025tacos} score of 57.4, the highest among the listed open-source models. It outperforms the time-aware TimeChat-Captioner by 10.2 points and also surpasses the proprietary Gemini-3.1-Pro and Gemini-2.5-Pro by 7.8 and 6.0 points, respectively. AudioMap further outperforms the larger Qwen3-Omni-Captioner by 15.9 points. These results show that, at 7B scale, AudioMap provides substantially stronger temporal grounding than both general-purpose Gemini baselines and caption-specialized open-source models in this comparison.

\paragraph{User Study.} 
We conduct a user study with 23 participants, who rate the generated captions on a 1--5 mean opinion score (MOS) scale across four dimensions: event coverage, fine-grained detail, correctness, and temporal grounding (Figure~\ref{fig:user_study}). AudioMap-7B achieves the highest mean score on all four dimensions: 4.56, 4.75, 4.38, and 4.41, respectively. Compared with the proprietary Gemini-2.5-Pro, these scores represent gains of 0.22, 0.69, 0.19, and 0.22 points. Together with the narrow 95\% confidence intervals, these consistently high ratings show that AudioMap delivers strong and reliable caption quality across all four evaluation dimensions.

\begin{figure*}[t]
    \centering
    \includegraphics[width=1\linewidth]{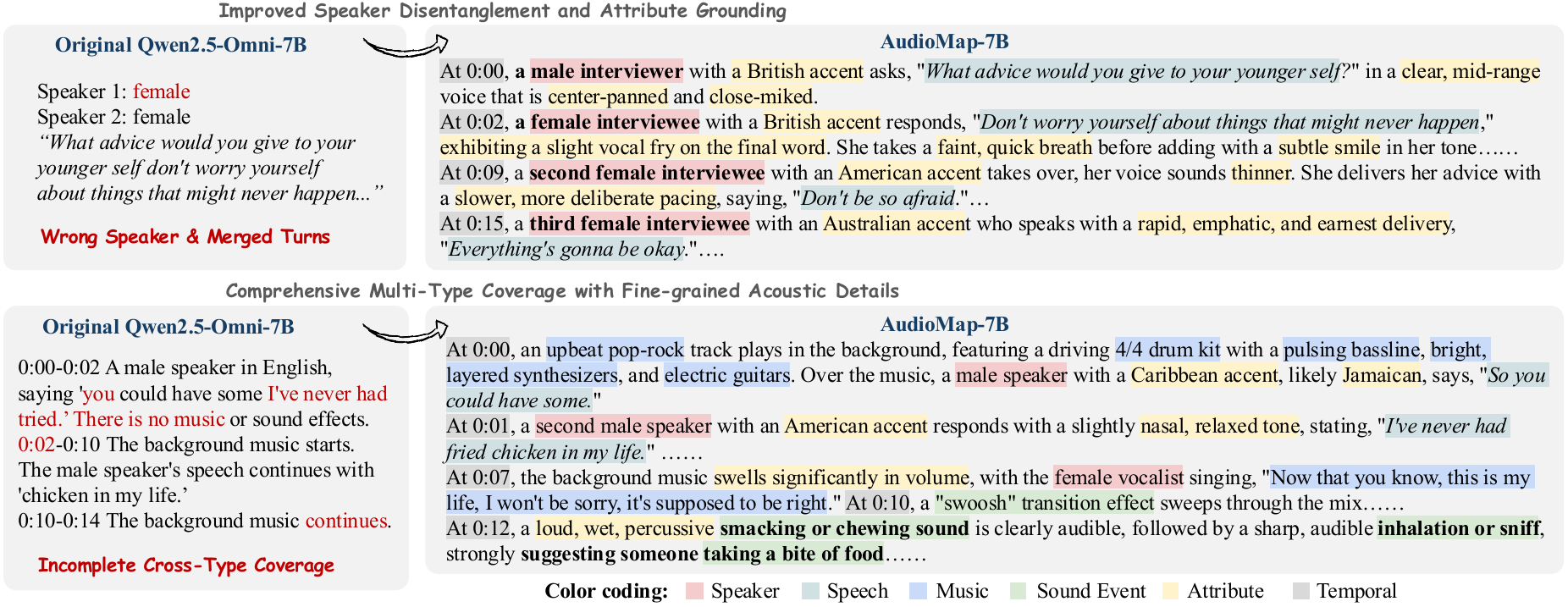}
    \vspace{-1.5em}
    \caption{Qualitative comparison of captioning capabilities between Qwen2.5-Omni-7B and AudioMap-7B.}
    \label{fig:qualitative_example}
    \vspace{-0.8em}
\end{figure*}

\begin{table}[t]
\centering
\caption{Ablation of the proposed reward terms. The best result in each metric is highlighted in bold.}
\tabcolsep=3.0pt
\vspace{-0.5em}
\label{tab:reward_ablation}
\resizebox{\linewidth}{!}{
\begin{tabular}{ccc@{\hspace{0.8em}}cccc}
\toprule
\multicolumn{3}{c}{\textbf{Reward terms}} & \multicolumn{4}{c}{\textbf{Evaluation}} \\
\cmidrule(lr){1-3}\cmidrule(lr){4-7}
$R_{\mathrm{ESR}}$ & $R_{\mathrm{ECTR}}$ & $R_{\mathrm{len}}$ & \textbf{Omni-Cloze} $\uparrow$ & \textbf{MMAR} $\uparrow$ & \textbf{MMAU} $\uparrow$ & \textbf{TACOS} $\uparrow$ \\
\midrule
\multicolumn{3}{l}{AudioMap-7B-SFT} & 59.7 & 60.6 & 70.6 & 54.6 \\
\midrule
\addlinespace[2pt]
\multicolumn{7}{l}{\textbf{\textit{GRPO reward configurations}}} \\
\checkmark & -- & -- & 62.7 & 61.9 & 71.7 & 54.6 \\
\checkmark & \checkmark & -- & \textbf{64.6} & 62.2 & \textbf{72.5} & 57.3 \\
\checkmark & \checkmark & \checkmark & \textbf{64.6} & \textbf{63.3} & 72.4 & \textbf{57.4} \\
\bottomrule
\end{tabular}
}
\vspace{-0.5em}
\end{table}

\paragraph{Qualitative Analysis.} 
Figure~\ref{fig:qualitative_example} presents representative comparisons between AudioMap-7B and the original Qwen2.5-Omni-7B. AudioMap better disentangles multiple speakers and overlapping acoustic events, while providing more comprehensive coverage of speech, music, and sound events with richer fine-grained attributes and clearer temporal organization. These examples demonstrate AudioMap’s stronger ability to produce complete, detailed, and time-aware audio descriptions.

\subsection{Ablation Studies}
\paragraph{Ablation on the Post-Training Pipeline.} 
Table~\ref{tab:pipe_ablation} shows how the progressive post-training pipeline builds AudioMap's final semantic and temporal capabilities. 
SFT establishes strong task-specific performance, while GRPO further improves both semantic and temporal metrics.
Removing asymmetric scoring consistently degrades performance, validating the stronger penalty for incorrect evidence. Removing the curriculum particularly hurts MMAU and TACOS, suggesting that curriculum learning mainly benefits multi-task understanding and temporal grounding.

\paragraph{Ablation on the Proposed Rewards.} 

Table~\ref{tab:reward_ablation} shows complementary effects of the proposed rewards. ESR primarily improves semantic metrics, raising Omni-Cloze, MMAR, and MMAU while leaving TACOS unchanged. Adding ECTR yields the largest temporal gain (+2.7 TACOS) together with further semantic improvements.  Adding length regularization changes the reported scores only modestly. Overall, ESR drives semantic coverage, while ECTR mainly improves temporal grounding.

\begin{table}[t]
\centering
\caption{Ablation study on different judge models. Best performances are highlighted in bold.}
\label{result_judge_model}
\tabcolsep=2.0pt
\vspace{-0.5em}
\resizebox{\linewidth}{!}{
    \begin{tabular}{lccccc}
    \toprule
    \textbf{Model}& \textbf{Omni-Cloze} $\uparrow$ & \textbf{MMSU} $\uparrow$ & \textbf{MMAR} $\uparrow$ & \textbf{MMAU} $\uparrow$ & \textbf{TACOS} $\uparrow$ \\
    \midrule 
    Qwen3-8B & 62.7 & 69.8 & 62.1 & 71.5 & 54.1 \\
    Qwen3.5-27B & 64.4 & \textbf{71.1} & 63.1 & 71.2 & 56.4 \\
    Qwen3.6-27B& \textbf{64.6} & 70.2 & \textbf{63.3} & \textbf{72.4} & 57.4 \\
    Gemma-3-27B-IT & 63.4 & 70.7 & 62.5 & 71.1 & \textbf{58.1} \\
    Mistral-Small-3.2-24B & 63.9  & 71.0  & 62.0 & 71.7 & 56.7 \\
    \bottomrule
    \end{tabular}
}
\vspace{-1.3em}
\end{table}

\paragraph{Reward Variants.}
Table~\ref{tab:pipe_ablation} further compares the proposed cloze-and-choice reward with holistic LLM-judge and checklist-based rewards. AudioMap outperforms the LLM-judge variant by 5.9 points on Omni-Cloze, 0.4 on MMAR, and 2.2 on MMAU. It also exceeds the checklist variant by 3.5, 2.1, and 1.0 points on these three metrics, respectively, and by 2.2 points on TACOS. The LLM-judge variant is 0.7 points higher on TACOS. This trade-off highlights a limitation of holistic scalar feedback: a single scalar signal can be sensitive to the judge's preferred wording or timestamp format and does not explicitly identify which acoustic details are missing or incorrect. In contrast, the cloze-and-choice formulation evaluates explicit semantic evidence and event-time pairs, yielding stronger results on Omni-Cloze and all three QA-based metrics while remaining competitive on TACOS.

\paragraph{Ablation across Judge Models.}

We further investigate the robustness of AudioMap to the choice of judge model during GRPO (Table~\ref{result_judge_model}). The 24–27B examiners from Qwen, Gemma, and Mistral yield comparable performance across benchmarks. Qwen3.6-27B achieves the strongest overall results, while Qwen3.5-27B and Gemma-3-27B-IT~\cite{gemma3} attain the best performance on MMSU and TACOS, respectively. In contrast, the smaller Qwen3-8B shows a more noticeable degradation, particularly on TACOS. These results suggest that the proposed reward is robust across examiner families, while sufficient examiner capacity remains important for reliable fine-grained and event-conditioned temporal supervision.
\section{Conclusion}
We present AudioMap, an RL framework for time-aware dense audio captioning based on a unified cloze-and-choice reward paradigm. ESR promotes faithful fine-grained evidence coverage, while ECTR explicitly aligns event semantics with timestamps. We further construct AudioMapCap-44K, a large-scale dataset for this task. AudioMap achieves SOTA performance among open-source models across diverse semantic and temporal benchmarks.


% \section*{Acknowledgments}

\bibliography{aaai2027}


% \clearpage

% \clearpage
\appendix
\section*{Appendix Overview}
This appendix is organized into the following two parts:
\begin{itemize}
    \item \textbf{A. AudioMapCap-44K:} Dataset construction, summary statistics, and quality-control procedure.
    \item \textbf{B. Experimental Setup:} Implementation details, training configurations, and evaluated baselines.
    % \item \textbf{C. Prompt Templates:} Captioning instruction pool and frozen-examiner prompts used for ESR and ECTR.
\end{itemize}

\section{A. AudioMapCap-44K}\label{appendix_data}
\subsection{AudioMapCap-44K Statistics and Quality Control}\label{appendix_dataset_stats}
After detailed annotation and quality screening, the final AudioMapCap-44K dataset contains 43,870 caption pairs spanning 769.70 hours. Clip durations range from 5.132 to 180.256 seconds, with a mean of 63.162 seconds and a median of 56.401 seconds. Table~\ref{tab:audiomapcap_composition} reports the audio-type and lyric distributions. Table~\ref{tab:audiomapcap_duration} reports the duration distribution.

\paragraph{Caption annotation.}
We select a subset of ASID-1M~\cite{li2026towards} and use Gemini-3.1-Pro~\cite{googledeepmind2026gemini31pro} to annotate each selected clip through an iterative three-stage pipeline. First, the model generates an initial caption that describes acoustic events, temporal boundaries, and event-level attributes such as timbre, emotion, and background noise. Second, the pipeline examines the initial caption for missing or underspecified information and conducts multiple rounds of targeted questioning. The model answers each question using the source audio and video as evidence. The visual stream is used at this stage as auxiliary cross-modal evidence to verify or disambiguate acoustic events that are difficult to resolve from audio alone.
Third, the pipeline merges the initial caption with the information collected from these multi-turn responses to produce a cohesive, time-aware dense caption.

\begin{table}[t]
\centering
\caption{Audio-type composition of AudioMapCap-44K. Marginal and exclusive shares use all 43,870 clips as the denominator. Lyric shares use the 31,442 clips containing music as the denominator.}
\label{tab:audiomapcap_composition}
\resizebox{\linewidth}{!}{
\begin{tabular}{lrr}
\toprule
\textbf{Category} & \textbf{Count} & \textbf{Share (\%)} \\
\midrule
\multicolumn{3}{l}{\textbf{\textit{Marginal audio types}}} \\
Speech & 35,773 & 81.5 \\
Sound effects & 31,791 & 72.5 \\
Music & 31,442 & 71.7 \\
Music with lyrics & 8,721 & 19.9 \\
\midrule
\multicolumn{3}{l}{\textbf{\textit{Mutually exclusive combinations}}} \\
Speech only & 2,357 & 5.4 \\
Sound effects only & 825 & 1.9 \\
Music only & 4,657 & 10.6 \\
Speech + sound effects & 9,246 & 21.1 \\
Speech + music & 5,065 & 11.5 \\
Sound effects + music & 2,615 & 6.0 \\
Speech + sound effects + music & 19,105 & 43.5 \\
\midrule
\multicolumn{3}{l}{\textbf{\textit{Lyrics among music-containing clips}}} \\
With lyrics & 8,721 & 27.7 \\
Without lyrics & 22,721 & 72.3 \\
\bottomrule
\end{tabular}
}
\end{table}

\begin{table}[t]
\centering
\caption{Clip-duration distribution of AudioMapCap-44K in 15-second intervals. Shares use all 43,870 clips as the denominator.}
\label{tab:audiomapcap_duration}
\resizebox{\linewidth}{!}{
\begin{tabular}{lrr@{\hspace{0.9em}}lrr}
\toprule
\textbf{Duration (s)} & \textbf{Count} & \textbf{Share (\%)} & \textbf{Duration (s)} & \textbf{Count} & \textbf{Share (\%)} \\
\midrule
0--15 & 4,826 & 11.0 & 105--120 & 4,326 & 9.9 \\
15--30 & 8,951 & 20.4 & 120--135 & 1,459 & 3.3 \\
30--45 & 3,810 & 8.7 & 135--150 & 1,241 & 2.8 \\
45--60 & 9,207 & 21.0 & 150--165 & 1,163 & 2.7 \\
60--75 & 2,192 & 5.0 & 165--180 & 1,104 & 2.5 \\
75--90 & 2,473 & 5.6 & $>180$ & 24 & 0.1 \\
90--105 & 3,094 & 7.1 & -- & -- & -- \\
\bottomrule
\end{tabular}
}
\end{table}

\paragraph{Quality screening.}
After caption construction, we apply two automatic checks to the full candidate set. First, rules remove captions that are too short, too long, or highly repetitive. Second, GPT-4.1~\cite{achiam2023gpt} evaluates the completeness of each remaining caption and assigns a score from 1 to 5. We retain samples scoring 4 or 5. We then manually inspect a subset of the retained samples against the source audio for event coverage, temporal accuracy, and hallucinated events. This process combines full-set automatic screening with targeted human review, providing scalable quality control for constructing complete and time-aligned SFT pairs.

\section{B. Experimental Setup}\label{appendix_setup}
\subsection{Implementation Details}
\paragraph{Training setup.}
We initialize AudioMap from Qwen2.5-Omni~\cite{xu2025qwen2_5omni} and train it on eight NVIDIA A800-SXM4 GPUs with 80\,GB of memory each. We use a single model and training pipeline for both input settings rather than training separate audio-only and audio-visual variants. Throughout SFT and GRPO, the vision encoder remains frozen, while the modality aligner, LLM backbone, and language modeling head are trainable. SFT uses all 43,870 examples in AudioMapCap-44K, with 80\% audio-only inputs and 20\% paired audio-visual inputs. GRPO uses 12,500 examples randomly sampled from AudioMapCap-44K and takes audio-only inputs. We train each stage of both the SFT and GRPO curricula for one epoch.

\paragraph{Supervised fine-tuning.}
For short clips, the learning rates are $1\times10^{-5}$ for the backbone, $5\times10^{-5}$ for the language modeling head, and $1\times10^{-4}$ for the aligner. We use a higher learning rate for the language modeling head than for the backbone because this head directly produces token predictions, allowing it to adapt more quickly to the structured dense-caption format. For long clips, we reduce the learning rates to $5\times10^{-6}$, $2\times10^{-5}$, and $2\times10^{-5}$, respectively.

\paragraph{Two-stage GRPO.}
The 12,500 GRPO examples are divided into two curriculum stages, as summarized in Table~\ref{tab:grpo_setup}. Stage~1 uses 6,000 clips shorter than 45 seconds and is approximately balanced across single and mixed audio types. It optimizes the policy with ESR and length regularization. Stage~2 uses 6,500 clips, including a small set shorter than 30 seconds, a majority between 30 and 90 seconds, and some longer than 90 seconds. It jointly optimizes the policy with ESR, ECTR, and length regularization.

\begin{table}[t]
\centering
\caption{Configuration of the two-stage GRPO curriculum. Reward weights are ordered as $(\lambda_{\mathrm{ESR}},\lambda_{\mathrm{ECTR}},\lambda_{\mathrm{len}})$.}
\label{tab:grpo_setup}
\resizebox{\linewidth}{!}{
\begin{tabular}{lcc}
\toprule
\textbf{Setting} & \textbf{Stage 1} & \textbf{Stage 2} \\
\midrule
Training examples & 6,000 & 6,500 \\
Clip duration & $<45$ s & Mostly 30--90 s \\
Reward weights & $(1.0,\,0,\,0.2)$ & $(1.0,\,1.0,\,0.2)$ \\
Model learning rate & $2\times10^{-6}$ & $1.5\times10^{-6}$ \\
Aligner learning rate & $1\times10^{-5}$ & $7.5\times10^{-6}$ \\
\bottomrule
\end{tabular}
}
\end{table}

We use Qwen3.6-27B~\cite{qwen36_27b} as the frozen examiner for ESR and ECTR. Each input produces eight rollouts with a temperature of 0.8, top-$p$ of 0.95, and top-$k$ of 50. The per-device batch size is 1, the gradient accumulation step is 8, and the generation batch size is 128. We set the KL coefficient to 0.06, weight decay to 0.1, gradient clipping to 0.5, and warmup ratio to 0.03 with a cosine learning-rate schedule whose minimum rate is 0.1 of the initial value. The maximum prompt and completion lengths are 14,336 and 4,096 tokens, respectively.

\subsection{Baselines}
We compare AudioMap with proprietary and open-source models under the evaluation protocols reported in Table 2 and 3. We organize the baselines by model access and task specialization.

\paragraph{Proprietary models.}
The audio-only comparison includes GPT-4o Audio~\cite{hurst2024gpt}, Gemini-2.5-Flash~\cite{comanici2025gemini}, Gemini-2.5-Pro~\cite{comanici2025gemini}, Gemini-3.1-Pro~\cite{googledeepmind2026gemini31pro}, and Qwen3.5-Omni-Plus~\cite{team2026qwen35}. The audio-visual comparison uses the three Gemini models and Qwen3.5-Omni-Plus~\cite{team2026qwen35}. These API-based systems provide general-purpose reference points for dense captioning from audio and audio-visual inputs.

\paragraph{Open-source generalist models.}
The audio-only comparison includes Qwen2-Audio-7B~\cite{chu2024qwen2}, Qwen2.5-Omni at 3B and 7B scales~\cite{xu2025qwen2_5omni}, Kimi-Audio-7B~\cite{ding2025kimi}, MiDashengLM-7B~\cite{dinkel2025midashenglm}, Step-Audio-2-mini-8B~\cite{wu2025step}, and Audio Flamingo 3-7B~\cite{ghosh2026audio}. For audio-visual evaluation, we include both Qwen2.5-Omni scales and Qwen3-Omni-Instruct-30B-A3B~\cite{xu2025qwen3}.

\paragraph{Caption-specialized models.}
We compare with Omni-Captioner-7B~\cite{ma2025omni}, AVoCaDO-7B~\cite{chenavocado}, and TimeChat-Captioner-7B~\cite{yaotimechat}. Omni-Captioner targets detailed captioning, while AVoCaDO and TimeChat-Captioner are designed for audio-visual captioning. TimeChat-Captioner further emphasizes time-aware, structured descriptions. We also include the larger Qwen3-Omni-Captioner-30B-A3B~\cite{xu2025qwen3} in the audio-only comparison. TAC~\cite{kumar2026tac} is not included because neither its model weights nor its training data are publicly available.

\end{document}